\documentclass[12pt]{article}

\usepackage{scicite}
\usepackage[pdftex]{graphicx}
\usepackage{times}
\usepackage{amsfonts,amsmath,amssymb,hyperref,lineno}

\newenvironment{sciabstract}{%
\begin{quote} \bf}
{\end{quote}}

\title{Finding shortest and nearly shortest path nodes in large substantially incomplete networks}

\author
{Maksim Kitsak$^{1,2\ast}$, Alexander Ganin$^{3,4}$, Ahmed Elmokashfi$^{5}$,
Hongzhu Cui$^{6,7}$,\\  Daniel A.Eisenberg$^{8}$, David L. Alderson$^{8}$, Dmitry Korkin$^{6,9,10}$\& Igor Linkov$^{11}$\\
\\
\normalsize{$^{1}$ Faculty of Electrical Engineering, Mathematics and Computer Science},\\ \normalsize{Delft University of Technology, Delft, The Netherlands,}\\
\normalsize{$^{2}$ Network Science Institute, Northeastern University, Boston, MA, 02115,}\\
\normalsize{$^{3}$ Department of Systems and Information Engineering,} \\ \normalsize{University of Virginia, Charlottesville, VA, 22904, USA,}\\
\normalsize{$^{4}$ U.S. Army Engineer Research and Development Center, Contractor, Concord, MA, 01742, USA,}\\
\normalsize{$^{5}$ Simula Metropolitan Center for Digital Engineering, Oslo, Norway,}\\
\normalsize{$^{6}$ Bioinformatics and Computational Biology Program,} \\ \normalsize{Worcester Polytechnic Institute, Worcester, MA 01609, USA,}\\
\normalsize{$^{7}$ Institute for Genomic Medicine,}\\ \normalsize{Columbia University Medical Center, New York, New York,}\\
\normalsize{$^{8}$ Operations Research Department, Naval Postgraduate School, Monterey, CA, 93943, USA,}\\
\normalsize{$^{9}$ Data Science Program, Worcester Polytechnic Institute, Worcester, MA 01609, USA,}\\
\normalsize{$^{10}$ Computer Science Department, Worcester Polytechnic Institute, Worcester, MA 01609, USA,}\\
\normalsize{$^{11}$ U.S. Army Engineer Research and Development Center,}\\ \normalsize{Environmental Laboratory, Concord, MA, 01742, USA,}\\
\normalsize{$^\ast$To whom correspondence should be addressed; E-mail:  m.a.kitsak@tudelft.nl}
}

\date{}

\begin{document} 


\baselineskip24pt


\maketitle


{\bf One sentence summary:} Shortest paths in real networks are geometrically localized and can be identified with network embedding methods even if networks are substantially incomplete.

\newpage
\begin{sciabstract}
Dynamic processes on networks, be it information transfer in the Internet, contagious spreading in a social network, or neural signaling, take place along shortest or nearly shortest paths. Unfortunately, our maps of most large networks are substantially incomplete due to either the highly dynamic nature of networks, or high cost of network measurements, or both, rendering traditional path finding methods inefficient. We find that shortest paths in large real networks, such as the network of protein-protein interactions (PPI) and the Internet at the autonomous system (AS) level, are not random but are organized according to latent-geometric rules. If nodes of these networks are mapped to points in latent hyperbolic spaces, shortest paths in them align along geodesic curves connecting endpoint nodes. We find that this alignment is sufficiently strong to allow for the identification of shortest path nodes even in the case of substantially incomplete networks. We demonstrate the utility of latent-geometric path finding in problems of cellular pathway reconstruction and communication security. 
\end{sciabstract}


\section*{Introduction}

Being the tallest building in the Western Hemisphere, the One World Trade Center (OWTC) is easily observable from virtually any point of the lower Manhattan island. Tourists can find their way to the building without a map as long as it stays in their line of sight. Another tourist attraction is the Peace Maze in Northern Ireland. The maze's exit is placed in the center and can also be spotted from anywhere within the maze. Nevertheless, finding the exit path from the maze is not straightforward. From the graph theory perspective, both the road system of Manhattan and the Peace Maze are graphs or networks, and both problems reduce to finding the shortest path connecting the origin with the destination. What makes New York City's most densely populated borough navigable is the geometric grid-like structure of its intersections. The line of sight to the OWTC is nothing else but the geodesic curve connecting the tourist's current location to its destination point, and the tourist may find her way to the OWTC by taking the streets with minimal deviation from this geodesic.

Our main finding is that  shortest paths in many real networks display geometric localization and can be identified using the same idea as in the One World Trade Center example. However, one crucial difference from the Manhattan road network is that links in many real networks, such as the Internet, social networks, and networks of molecular interactions, are not determined by \emph{physical} proximities of their nodes. On the contrary, these networks are characterized by \emph{effective} geometries, which are often referred to as latent or hidden.  Nodes in these networks can be mapped to points in latent spaces by an optimization procedure, often called network embedding, such that in the resulting map, network links are likely to connect nodes separated by small distances in the latent space~\cite{Goyal2018graph,Cui2019survey}.

\section*{Latent-geometric organization of networks and shortest paths in them}

Recent works indicate that common topological properties of real networks, such as the hierarchical organization, the heterogeneity in the number of connections per node, strong clustering coefficient, and self-similarity~\cite{boguna2020network}, are best mapped into latent spaces, which are hyperbolic rather than {\it Euclidean}. Notable examples of real networks with effective hyperbolic geometries are the PPI networks~\cite{lobato2018latent}, the Internet~\cite{Boguna2010sustaining}, and human proximity networks~\cite{rodriguezflores2020hyperbolic}.  At the same time, there is no consensus on the extent to which shortest paths in these networks align along geodesic curves. While an earlier work demonstrated that hyperbolic geometry could be used to find Internet routing paths~\cite{Boguna2010sustaining}, more recent work finds that the topologically shortest paths are statistically different from geometrically shortest paths~\cite{cannistraci2020geometrical}.

Building on these works, we study the prospects of shortest path finding using the hyperbolic representations of the AS-level Internet and the human similarity-based PPI network. In the AS-level Internet, nodes are Autonomous Systems (ASes), and connections between them are contractual agreements governing data flows between ASes, SM section SV. The similarity-based PPI network is the derivative of the traditional PPI network. In our construction, two proteins are linked if they have a statistically significant number of common interaction protein partners, see SM section SVI. Since two proteins with common interaction partners can be interpreted as similar, we refer to the resulting network as the similarity-based PPI network. We discuss the basic ideas of hyperbolic network mapping in Methods and preserve the associated technical details for SM section SIV.

\section*{Distance to  geodesic and path finding accuracy}

To demonstrate the main finding, we first visualize two shortest paths of the AS-level Internet in its 2-dimensional hyperbolic representation.  Figure~1{\bf a} demonstrates that nodes comprising the two shortest paths are not random but tend to lie in the geometric vicinity of corresponding hyperbolic geodesics connecting the endpoints of the two shortest paths.
To quantify the observed alignment, we measured distances from the shortest path nodes to the geodesic curve, Fig.~1{\bf b}. To do so, we employed an approximation for a distance from point $C$ to the hyperbolic geodesic curve $\gamma(A,B)$ connecting points $A$ and $B$ in hyperbolic disk $\mathbb{H}^{2}$:
\begin{equation}
d(C,\gamma(A,B)) = \frac{1}{2}\left[d(A,C) + d(B,C) - d(A,B)\right] + \ln 2,
\label{eq:distABC}
\end{equation}
where $d(X,Y)$ is the distance between points $X$ and $Y$ in $\mathbb{H}^{2}$, see Methods.

We calculated distances from all network nodes to the hyperbolic geodesic connecting the  AS5392-AS8875 node pair, finding that all 6 shortest path nodes are among the 12 closest to the geodesic nodes, see Fig.~1{\bf c}. Although not part of the original shortest path, the remaining 6 closest to the geodesic nodes may form alternative shortest paths if the Internet topology is perturbed. To verify this claim, we calculated for each node its \emph{path relevance}, which we defined as the probability to be on the shortest path in case  network links are removed uniformly at random with probability $q=0.5$. We found, see Fig.~1{\bf d}, that node path relevance is anti-correlated with the distance to geodesic, and the closest to the geodesic nodes are characterized by the largest path relevance values, confirming our claim. The anti-correlation between path relevance and the distance to geodesic  also indicates that distance to geodesic is capable of identifying not only the shortest path nodes but also nodes that may belong to a  shortest path if the network topology is perturbed.

Before further explaining our results, we need to discuss some technical challenges and provide some rigorous definitions. First, many real networks are characterized by the small-world property: network-based distances between the nodes scale logarithmically~\cite{Watts1998} or even sub-logarithmically~\cite{cohen2003scale} as a function of network size $N$.  Similarly, the sets of the shortest path nodes are extremely small compared to the network size, making the shortest path classification problem extremely unbalanced~\cite{maimon2010data}. Second, from the perspective of dynamic processes, such as communication or viral spreading, the propagation along nearly shortest paths is not much worse than the propagation along the shortest paths. These two observations motivated us to consider nearly shortest path nodes along with the shortest path nodes. To do so, we combined both entities under the umbrella of the path relevance metric, Fig.~1{\bf d}, and defined nearly shortest path nodes as nodes with path relevance values exceeding $0.05$.  Nearly shortest path nodes address both challenges. Indeed, they contain not only the original shortest paths but also nodes on slightly less optimal paths. Also, the sets of the nearly shortest path nodes are larger then those of the original shortest paths, reducing the classification imbalance.

To quantify the accuracy of the identification of nearly shortest path nodes, we use two metrics. The first metric is the statistical precision score, defined as the fraction
of  true nearly shortest path nodes. The second metric is the number of node removals necessary to disrupt all paths between the pair of nodes of interest, SM section SVII. While indirect, the second metric provides an insight into how effective a path identification method of interest is in finding possible path deviations.

We compare the accuracy of the distance to geodesic metric  $d(C,\gamma(A,B))$ to its network-based counterpart,  $d_{\rm nb}(C|A,B) = \ell_{A,C} + \ell_{A,B}$, which is the sum of the network-based distances from node $C$ to path endpoints $A$ and $B$. Since shortest paths are known to traverse the most connected nodes in networks~\cite{NEWMAN2010}, it is tempting to use node degree as an alternative path relevance metric. The node degree is, however, of limited practical use since it is usually hard, if not impossible, to control large-degree nodes. The most connected Internet ASes are the largest Internet Service or Content Providers that are rarely affected by adverse events~\cite{green2018leveraging}. Proteins with large numbers of interaction partners are hard to manipulate due to the increased possibility of side-effects~\cite{Barabasi2011}. Large degree nodes are usually shared by many communication paths, as quantified by the betweenness centrality~\cite{freeman1977set}, and manipulating them will affect not only the path of interest but also all other paths in the network.  As seen in Fig.~2{\bf a}, removing nodes with the largest degree values affects the lengths of all paths, decreasing the average inverse network diameter.

\section*{Finding nearly shortest paths in incomplete networks}

To quantify the alignment of nearly shortest paths, we conducted a series of path finding experiments on the AS Internet and the similarity-based human PPI networks with varying fractions of missing links. In all experiments, we first computed the true nearly shortest path nodes using the original network. We then removed randomly selected links and tried to identify nearly shortest path nodes on incomplete networks. In the case of the distance to  geodesic, we first obtained the hyperbolic maps of the incomplete networks. For every node pair $A$-$B$ of interest, we then determined corresponding hyperbolic geodesic $\gamma(A,B)$ and then used distance to geodesic $d(C,\gamma(A,B))$, Eq.~(\ref{eq:distABC}), to quantify the proximity of other network nodes to $\gamma(A,B)$. Finally, we computed the network-based scores $d_{\rm nb}(C|A,B)$  directly on incomplete networks.

We observe that the accuracy of the distance to geodesic only decreases mildly as the fraction of missing links increases, Fig.~2{\bf b} and Fig.~S12.  This result is in sharp contrast with the accuracy of the network-based method, which decreases fast as the rate of the missing links increases. While both the network-based method and the distance to geodesic yield comparable results on complete networks, in cases of substantially incomplete networks, $70\%$ and $90\%$ of the missing links, the precision of the distance to geodesic is approximately twice than that of the network-based method, Fig.~2{\bf b}. Similarly, in the case of the $70\%$ missing link rate, it takes six times more node removal steps, on average, to disrupt all paths  with the network-based ranking, compared to that of the distance to geodesic ranking,~Fig.~2{\bf b}. This is the case since network-based path finding methods are extremely sensitive to missing network data. Traditional shortest path finding methods, like the classical Dijkstra algorithm~\cite{Dijkstra1959note}, iteratively explore network-based neighborhoods of the path end-nodes. As a result, network-based methods rely on the accuracy of the network data and are doomed to fail if some of the network data is missing. Latent-geometric maps of networks, on the other hand, provide an effective {\it mean-field} image of the network, which is not very sensitive to uniformly missing network data~\cite{kitsak2019link}. As a result, the distance to geodesic is reliable even in the case of substantially incomplete networks, where the number of missing links exceeds that of known links.

\section*{Identifiability limits for nearly shortest paths}

To explore the limits of the distance to geodesic metric in the identification of shortest path nodes, we conducted experiments on incomplete random hyperbolic graph (RHG) models~\cite{Krioukov2010hyperbolic,Aldecoa2015}. RHG models are obtained by sprinkling network nodes into a 2-dimensional hyperbolic disk $\mathbb{H}^{2}$ and connecting node pairs with distance-dependent probabilities,  SM section~SIII. RHGs are used as null models in hyperbolic network embedding methods and also allow the generation of synthetic networks with scale-free degree distribution $P(k) \sim k^{-\lambda}$ with variable exponent $\lambda \in (2,3)$ and variable degree of geometricity that is controlled by the temperature parameter $T \in [0,1]$. In the limiting case of $T=0$, network links
are only allowed between node pairs at small latent distances in $\mathbb{H}^{2}$. In this case, network geometricity is the strongest since all network links are short-range. As $T$ increases, connections at larger distances are allowed with increasing probabilities, leading to weaker network geometricity, SM section SVII. 

To identify paths in RHGs using distance to geodesic, we first erased their original node coordinates in $\mathbb{H}^{2}$ and then re-learned them using the HL embedder~\cite{kitsak2019link}. Our path finding experiments for incomplete RHGs  suggest that the accuracy of the distance to geodesic is nearly independent of the degree distribution exponent $\lambda$ and strongly depends on network geometricity, as quantified by temperature $T$, Fig.~2{\bf c,d}.  As $T$ increases, the latent-geometric path finding accuracy decreases and becomes comparable to that of the network-based method, Fig.~2{\bf c,d} and Figs.~S8{\bf a}, S9{\bf a}, S10{\bf a}, S11{\bf a}. For comparison, we obtained similar heatmaps for the network-based distance, Figs.~S8{\bf b}, S9{\bf b}, S10{\bf b}, S11{\bf b}, observing the distance to geodesic is superior to the network-based distance nearly in the entire range of $\lambda$-$T$ parameters, except for the largest $T$ values, as shown by the dashed lines in Fig.~2{\bf c,d}.

Distance to geodesic can be invaluable in the analysis of paths in incomplete networks. One family of applications, to this end, concerns the validation of existing communication paths. The validation of communication paths is much needed in distributed communication networks, such as the Internet. Another family of applications stems from the ability of distance to geodesic to find alternative nearly shortest paths. The latter task of finding possible path deviations is relevant not only in communication networks but also in cellular pathways, as we discuss below.

\section*{Assessing the integrity of routing paths with distance to geodesic}

Autonomous Systems (ASes) comprising the Internet are independent organizations. Hence, the information on how to reach devices in another AS is not readily available to them. This reachability information is disseminated by the Border Gateway Protocol (BGP)~\cite{rekhter2014rfc}. BGP belongs to the family of path-vector routing protocols: ASes share with their neighbor ASes paths to various destinations known to them. BGP routing is based on trust: ASes accept routing paths advertised by their neighbors without strict integrity tests, Fig.~3{\bf a}. With the increasing number of recent cyberattacks and routing instabilities, path integrity checks are becoming much desired~\cite{sermpezis2018survey}. One class of cyberattacks is the BGP prefix hijacking. During a BGP prefix hijack, an AS either claims ownership of IP address prefixes that are owned by other ASes or falsely announces that it can provide transit to a prefix or a set of prefixes. This attack can compromise the affected data flows by either exposing them to malicious actors or simply misrouting them~\cite{sermpezis2018survey}, Fig.~3{\bf a}.

While interdomain Internet routing paths are known to be longer than shortest paths due to many factors, including AS business relationships~\cite{gao2001inferring}, the extent of their inflation, as found in Ref.~\cite{gao2002extent}, places them into the category of nearly shortest paths. Since the latter tend to align along geodesics in the Internet hyperbolic representation, we propose that the integrity of Internet routing paths can be assessed by charting them in the latent hyperbolic space. We expect genuine routing paths to localize in the vicinity of the hyperbolic geodesic connecting path endpoints, Fig.~3{\bf b}. Fake routing paths, on the other hand, are expected to significantly deviate from corresponding geodesics, allowing for their detection, Fig.~3{\bf c}. 

To quantify the alignment of routing paths, we propose the hyperbolic stretch $D_{\Omega(A,B)}$, which we define as the maximum normalized distance from path $\Omega(A,B)$ to the geodesic curve $\gamma(A,B)$ connecting path endpoints, $D_{\Omega(A,B)} = {\rm max}_{C \in \Omega(A,B)}  \frac{d(C, \gamma(A,B))}{d(A,B)}$, SM section~SV. We studied three recent BGP instability events involving MainOne Telecom in Nigeria, Malaysia Telecom (MT),  and Rostelecom (PJSC), respectively, SM section~SV. We analyzed routing paths involving the affected ASes that were advertised both before and during each instability event. As seen from Fig.~3{\bf d} and Fig.~S4,  stretches of routing paths advertised during the hijack events are significantly larger than those before the event, suggesting that ASes may use the hyperbolic stretch to reject questionable routing paths, SM section~SV. In contrast to network-based methods, ASes do not need to know the exact AS-level network to assess routing path integrity. To compute the hyperbolic stretch, one only needs to know the hyperbolic coordinates of ASes constituting the routing path of interest. Our results indicate that AS coordinates can be computed with sufficient accuracy even  when a large fraction of network links is unknown.

\section*{Exploring the neighborhoods of cellular pathways in human PPI network}

Complex cellular processes and many diseases involve multitudes of genes and their products that are organized into molecular pathways. It is also not uncommon for the same pathway to be linked to several diseases or be intrinsically involved in more than one molecular process. One of the key questions arising from the point of view of network biology is the functional relationship of the pathway genes with the genes located in its proximity. 

Are cellular pathways akin to communication paths? There is no clear-cut answer to this question: different from the communication paths, cellular pathways often have no single origin and destination. Neither do cellular pathways conduct traffic. While the communication paths have a clear objective to be optimal, there is no such a requirement for cellular pathways, although there is an expectation that the cellular pathways evolved to become optimal~\cite{nam2011role}. In the light of these differences, an intriguing finding of our work is that some cellular pathways align along the geodesic curves when drawn on the hyperbolic representations of PPI networks, see Figs.~4{\bf a} and Fig.~S6.  

We studied the ubiquitin proteasome (UPP), the transforming growth factor beta (TGFb), and the cell cycle pathways. For each pathway, we identified its localizing hyperbolic geodesic curve using the least squares fitting for pathway proteins, SM section SVI. We have observed that  the two halves of the geodesic curves in these pathways often naturally split the proteins associated with the pathways to the functionally related subsets of proteins. For the UPP pathway, Fig.~4{\bf a}, we found that proteins in the geometric vicinity of the fitted geodesic are associated with E2, and E3 enzyme classes. We found that the larger geodesic branch 10 o'clock is associated with  E2 and E3 classes but not with E1. The 12 o'clock branch, on the other hand, is exclusively associated with the E3 class, Fig.~4{\bf a}.

To further explore the analogy between cellular pathways and communication paths, we asked  if other genes that are functionally similar to the ones in the geodesic but lie outside of it can be found using their  proximity to the geodesic. To answer this question, we considered $100$ genes in the latent-geometric proximity to the ubiquitin-proteasome pathway (UPP), as quantified by the distance to the fitted geodesic, Eq.~(\ref{eq:distABC}). We used DAVID tools~\cite{huang2009systematic}, a bioinformatics framework specifically designed to provide systematic functional analysis for a large group of genes, to functionally cluster groups of the proximal genes independently from the genes in the geodesic cluster, finding $6$ major  clusters, SM section SVI.  As expected, the most highly ranked cluster contained the terms related to ubiquitination, cluster $1$, Fig.~S7{\bf a} and Table~S1. Interestingly, the other clusters include genes associated with the immune response signaling pathway, cluster $2$, Fig.~S7{\bf b}, viral infections, clusters $3$, Fig.~S7{\bf c} and $4$, Fig.~S7{\bf d}, pathways associated with several types of cancer, clusters $3$ and $4$, as well as zinc fingers, cluster $5$, Fig.~S7{\bf e}, and DNA repair, cluster $6$, Fig.~S7{\bf f} and Table~S1. Each of the functional groups is naturally associated with ubiquitination~\cite{hu2016ubiquitin,deng2020role,gustin2011viral,geng2010rad,morrow2015targeting}. Indeed, ubiquitination is a key mechanism regulating signal transduction and mediating both innate and adaptive immune responses~\cite{hu2016ubiquitin}. On the other hand, the principal role of protein containing zinc finger domains in ubiquitination has emerged only recently~\cite{bienko2005ubiquitine,peisley2014structural,ali2012novel,rahighi2009specific}.

Our observations indicate that the UPP, TGFb, and cell cycle pathways are organized similarly to communication paths. Not only are these pathways aligned along geodesic curves, but other genes in the latent geometric vicinity to them appear to be functionally related. To complete the analogy, it is tempting to ask whether perturbations in protein interaction networks result in perturbed pathways, which follow the same geodesic. While we do not have the answer to this question yet, we hypothesize that two likely scenarios can happen when replacing a malfunctioning gene, e.g., due to a deleterious mutation. The first scenario is when the new gene is a parlor of the replaced gene, and the reason that it was not in the original pathway is that it performs the function less efficiently than the original gene. In this case, the new pathway will be longer compared to the original one. The second scenario is when the new gene corresponds to an alternatively spliced variant that could perform the function in the same  manner or even more efficiently but is under-expressed compared to the original gene that is the primary spliced variant. In this case, if the original gene is removed, the new node will become the new primary splice variant, with a much higher relative expression due to the absence of the old primary splicing variant. As a result, we expect the new pathways to be of the same length or even shorter.

\section*{Summary and discussion}

Taken together, we established that latent-geometric geodesics serve as fairways for shortest and nearly shortest paths nodes in geometric networks. Nodes in the vicinity of these geodesics are likely to lie on shortest paths or may become shortest path nodes if network the topology is perturbed. Our finding can  be either a curse or a blessing, depending on the circumstances. One could exploit the geometric localization of shortest paths to disrupt or eavesdrop on communication paths of interest. On the other hand, the knowledge of geodesic fairways may help identify alternative optimal paths and rule out inefficient or fraudulent paths in communication networks.

It is important to emphasize that there is no {\it one-size-fits-all} solution to the shortest path problem. In order to identify shortest path nodes in a partially known network, one needs to know both the mechanisms of network formation and the character of missing data. Distance to geodesic, in this respect, assumes that link formation in the network is captured by its latent geometry, and unknown links are missing uniformly at random. The first condition is a must: one cannot expect to identify shortest paths in non-geometric networks using geometric methods. The second assumption could probably be relaxed. While our embedding algorithm is designed for networks with uniformly missing links, it should be straightforward to generalize it to the special case when the probability of a missing link is also a function of a latent distance between the nodes. The accurate embedding of non-uniformly incomplete networks will further improve the accuracy of path inference tasks on real networks.

\section*{Materials and methods}

{\it Path relevance and nearly shortest path nodes.} We define the relevance of node $C$ to paths connecting nodes $A$ and $B$ as the probability that $C$ belongs to the shortest path connecting $A$ and $B$ if network links are removed uniformly at random with probability $q$. The path relevance metric allows to identify not only
the original shortest path nodes but also nodes that may become shortest either if the network topology is perturbed or if the original shortest path becomes unavailable. 
In all experiments, we set the missing link probability to $q=0.5$. We define \emph{nearly shortest path nodes} connecting nodes $A$ and $B$ as the nodes with the $A$-$B$ path relevance exceeding 0.05.

{\it Distance to geodesic.} Distance between two points $\{r_{i}, \theta_i\}$ and $\{r_{j}, \theta_j\}$ in the 2-dimensional hyperbolic disk $\mathbb{H}^{2}$ is given by the hyperbolic law of cosines
\begin{equation}
\cosh \zeta d_{ij} = \cosh \zeta r_i \cosh \zeta r_j - \sinh \zeta r_i \sinh
\zeta r_j \cos \Delta \theta_{ij},
\label{eq:hypercos}
\end{equation}
and for sufficiently large $r_{i}$ and $r_{j}$ values is closely approximated by $\mathbf{x}_{ij} = r_i + r_j + {2 \over \zeta} \ln \left( \sin (\Delta \theta_{ij} /2) \right)$, see SM section~SII.

Distance from point $C$ to geodesic $\gamma(A,B)$ as the shortest distance from $C$ to any point on $X \in \gamma(A,B)$:
\begin{eqnarray}
d(C,\gamma(A,B)) = {\rm min} ~~d(C,X),\\
{\rm s.t.} ~~X \in \gamma(A,B)
\end{eqnarray}
The distance to the hyperbolic geodesic $d(C,\gamma(A,B))$ is closely approximated by Eq.~(\ref{eq:distABC}), see SM section~SII.

{\it Network mapping} or embedding into a latent space $\mathcal{M}$ is a procedure of determining the coordinates of nodes constituting the network in this space. In this work we map AS Internet and the similarity-based protein interaction network to the 2-dimensional hyperbolic disk using the HL Embedder algorithm~\cite{kitsak2019link}. Similar to other hyperbolic embedders, HL embedder maps network  nodes to points $\{r_{i}, \theta_{i}\}$, $i=1,...,N$, in a hyperbolic disk $\mathbb{H}^{2}$ by maximizing the posterior probability $\mathcal{L} \left(  \{ r_i,\theta_i \}| a_{ij}\right)$ that the network with the adjacency matrix $a_{ij}$  has given node coordinates and is generated as the RHG model, SM section SIII.  By the Bayes' rule,  $\mathcal{L} \left(  \{ r_i,\theta_i \}| a_{ij}\right) \propto \frac{\mathcal{L}  \left(  a_{ij}| \{ r_i,\theta_i \}|\right)} {\rm Prob}\left(\{ r_i,\theta_i \}\right)$, where $\mathcal{L}  \left(  a_{ij}| \{ r_i, \theta_i \}|\right)$ is the likelihood that the network $a_{ij}$ is generated as the RHG, given node coordinates  $\{ r_i,\theta_i \}$, and the ${\rm Prob} \left( \{ r_i,\theta_i \} \right)$ is the prior probability of node coordinates generated by the RHG. Since links $\{ij\}$ in the RHG are established independently with probabilities depending on hyperbolic distances $\{\mathbf{x}_{ij}\}$ between the nodes, $\mathcal{L}\left(  a_{ij}| \{\mathbf{x}_{i}\}\right) = \prod_{i < j} \left[p\left(x_{ij}\right)\right]^{a_{ij}}  \left[1  - p\left(x_{ij}\right)\right]^{1 - a_{ij}}$. The HL embedder is freely available at the github repository~\cite{codehlembedder}.


\bibliographystyle{Science}
\bibliography{bib,ae_bib,hc_bib}

\section*{Acknowledgements}
The authors thank to I. Voitalov, H. Hartle, W.L.F. van der Hoorn, P.F.A. Van Mieghem, and D. Krioukov for stimulating discussions. This work was supported by the U.S. Defense Threat Reduction Agency. MK was additionally supported by NSF grant IIS-1741355, ARO grants W911NF-16-1-0391 and W911NF-17-1-0491, and the NExTWORx project. The views and opinions expressed in this article are those of the individual authors and not those of the U.S. Army, or other sponsor organizations.

\section*{Supplementary Material}

Material and Methods\\
Figs.~S1 to S12\\
Supplementary Data\\

\section*{Figure Captions}

\noindent {\bf Fig. 1.}  {\bf Latent geometry uncovers shortest and nearly shortest paths in the Internet at the Autonomous System level.}
{\bf a}, Hyperbolic map of the Internet at the Autonomous System (AS) level. See SM section~SV for data collection and hyperbolic mapping details. The latent space
is the 2-dimensional hyperbolic disk and each point corresponds to an Autonomous System. Yellow squares and pink circles highlight ASes corresponding to communication paths between AS5392-AS8875 and AS1224-AS11650 pairs. Shown are the nodes with path relevance exceeding 0.1, see Methods.
Sizes of squares and circles are larger for nodes with higher path relevance. {\bf b}, Schematic distance from point $C$ to geodesic $\gamma(A,B)$ drawn between points $A$ and $B$. {\bf c}, The distribution of distances to the $\gamma(AS5392,AS8875)$ geodesic from (light blue) shortest path nodes, (dark blue) nodes with path relevance larger than 0.05, and (red) all Internet nodes.

\noindent {\bf Fig. 2.}  {\bf The accuracy of nearly shortest path finding for the AS Internet} {\bf a}, The average inverse network-based distance between the AS Internet node pairs as a function of the number of removed nodes. The average inverse distance for each data point is estimated over 1,000 randomly selected node pairs. The average inverse distance serves as the measure of the \emph{collateral damage} to the network: the smaller is the value, the larger are the network-based distances between nodes. We consider three node removal strategies: (i) nodes with the largest degree, (ii) nodes with the smallest latent distance to $\gamma(AS5392, AS8875)$ geodesic, and (iii) nodes with the smallest network-based distances to the AS5392 - AS8875 pair. The largest degree strategy is the most invasive, resulting in the nearly two-fold decrease of the average inverse network-based distance upon removing $100$ largest degree nodes. This result is consistent with earlier fundings on attack tolerance of complex networks~\cite{albert2000error}.  {\bf b}, The accuracy of distance to geodesic in finding nearly shortest path nodes in the incomplete AS Internet network. The distance to geodesic strategy is juxtaposed against the network-based strategy, where path relevance of any node $C$ is quantified by the sum of network-based distances to path endpoints, $d_{\rm nb}(C|A,B) = \ell_{A,C} + \ell_{A,B}$. We evaluate the average precision scores and the average number of node removals needed to disrupt node pairs of interest. Each data bar is the average over 1,000 randomly selected node pairs, and error bars display one standard deviation. Note that the relative performance of the distance to geodesic increases as the fraction of missing links increases.  {\bf c, d}, The accuracy of distance to geodesic in finding nearly shortest paths in incomplete synthetic networks. Synthetic networks are constructed as RHGs of $N=5,000$ nodes, average degree $\langle k \rangle = 10$,  and variable degree distribution, $\lambda \in (2,3)$, and geometricity $T \in (0,1)$ parameters. After the construction of each network, we remove each of its links with probability $p=0.5$. The heatmaps consist of $9\times 9 = 81$ points, each point corresponding to the average of 100 randomly selected node pairs separated by the $\Delta \theta = \frac{\pi}{8}$ angle in $\mathbb{H}^{2}$. For technical details and other angles, see SM section SVII. Panel {\bf c} displays precision scores, while panel {\bf d} displays the number of node removals needed to disconnect all paths between the node pair of interest. Note that the path finding accuracy is nearly independent of degree distribution. The highest path finding accuracy is achieved at temperature $T=0$, when the geometricity of RHGs is the strongest and decreases as $T$ increases. Marked on panels {\bf c, d} are inferred parameters of real networks. These networks are AS Internet  ($\lambda = 2.1$, $T=0.7$) and similarity PPI ($\lambda=2.1$, $T=0.4$) that are studied in this work, the pretty-good-privacy (PGP) web of trust (PGP web of trust, $\lambda=2.1$, $T=0.6$) from Ref.~\cite{Papadopoulos2012popularity}, the network of human metabolic interactions (Human metabolism, $\lambda=2.55$, $T=0.65$), Ref.~\cite{Serrano2012uncovering}, and the network of protein-protein interactions (original PPI, $\lambda=2.65$, $T=0.7$), Ref.~\cite{lobato2018latent}.

\noindent {\bf Fig. 3.}  {\bf Anomaly detection in interdomain Internet routing} {\bf a}, Interdomain routing at a glance. Shown is the AS Internet toy network, where nodes are autonomous systems (ASes) and links are data transfer agreements. Shown on the right-hand side is the distributed calculations of paths to AS8. The generation of paths on the left-hand side is an example of routing instability. AS1 falsely claims the direct connection to AS8. This information  is propagated by the BGP routing protocol first to AS5 and the to AS8. As a result, AS5 and AS7 have fake path information to AS8. Routing paths are hard to verify using network-based methods since ASes do not have global AS Internet connectivity information. {\bf b,c,d,}
routing paths following the hijack of the Google AS (AS15169) prefixes by the MainOne AS (AS37282), see SM section SV. {\bf b}, An example of a routing path traversing Nigerian Internet Service Provider MainOne (AS37282) announced before the prefix hijack event. Green circles are ASes constituting the path; the solid red line is the hyperbolic geodesic connecting the origin-destination pair. Distance to geodesic for every node constituting the routing path is shown above the hyperbolic map. The hyperbolic stretch of the BGP path, Eq.~(\ref{eq:distABC}), is the largest normalized distance to geodesic from routing path nodes, $D_{\Omega(A, B)} = 0.23$. {\bf c}, An example of a routing path announced during the prefix hijack of MainOne AS.  MainOne announced a direct connection to Google AS (AS15169), leading to a large-scale cascade of false BGP path announcements. Shown is one of these false BGP paths originating at AS29140, traversing the MainOne AS (AS37282) and ending at the Google AS (AS15169). Distance to geodesic for every node constituting the routing path is shown above the hyperbolic map. The hyperbolic stretch of the routing path is $D_{\Omega(A, B) = 0.74}$, indicating that the BGP path is less conformal to the latent-geometric geodesic, compared to the one in panel {\bf b}. Red squares are ASes constituting the path; the solid red line is the hyperbolic geodesic connecting the origin-destination pair.  {\bf d}, Distributions of hyperbolic stretches for BGP paths announced (blue) before and (red) during the MainOne AS prefix hijack. Note that BGP paths announced before the hijack are characterized by significantly smaller hyperbolic stretch values than those announced during the hijack event.

\noindent {\bf Fig. 4.} {\bf Latent-geometric localization of the $UPP$ pathway.} {\bf a}, the hyperbolic map of the similarity-based human PPI network and the ubiquitin proteasome pathway. Pathway proteins are colored according to their functional groups, background nodes are non-pathway proteins comprising the network. E1, E2, and E3 are the three classes of enzymes associated with ubiquitination: E1s and E2s are ubiquitin-activating and ubiquitin-conjugating enzymes, respectively, while E3s correspond to ubiquitin-protein ligase. The solid line displays the hyperbolic geodesic fitting pathway proteins.   Panel {\bf b} depicts $6$ clusters of proteins in the latent-geometric vicinity of the UPP pathway geodesic. Proteins are colored based on the number of clusters they belong to. Black squares depict UPP pathway proteins. See also Fig.~S7, which depicts each of the 6 clusters separately.

\newpage
\clearpage
\centerline{\bf Latent geometry uncovers shortest and nearly shortest paths in the Internet at the Autonomous System level.}

\begin{figure}
\includegraphics[width=7in]{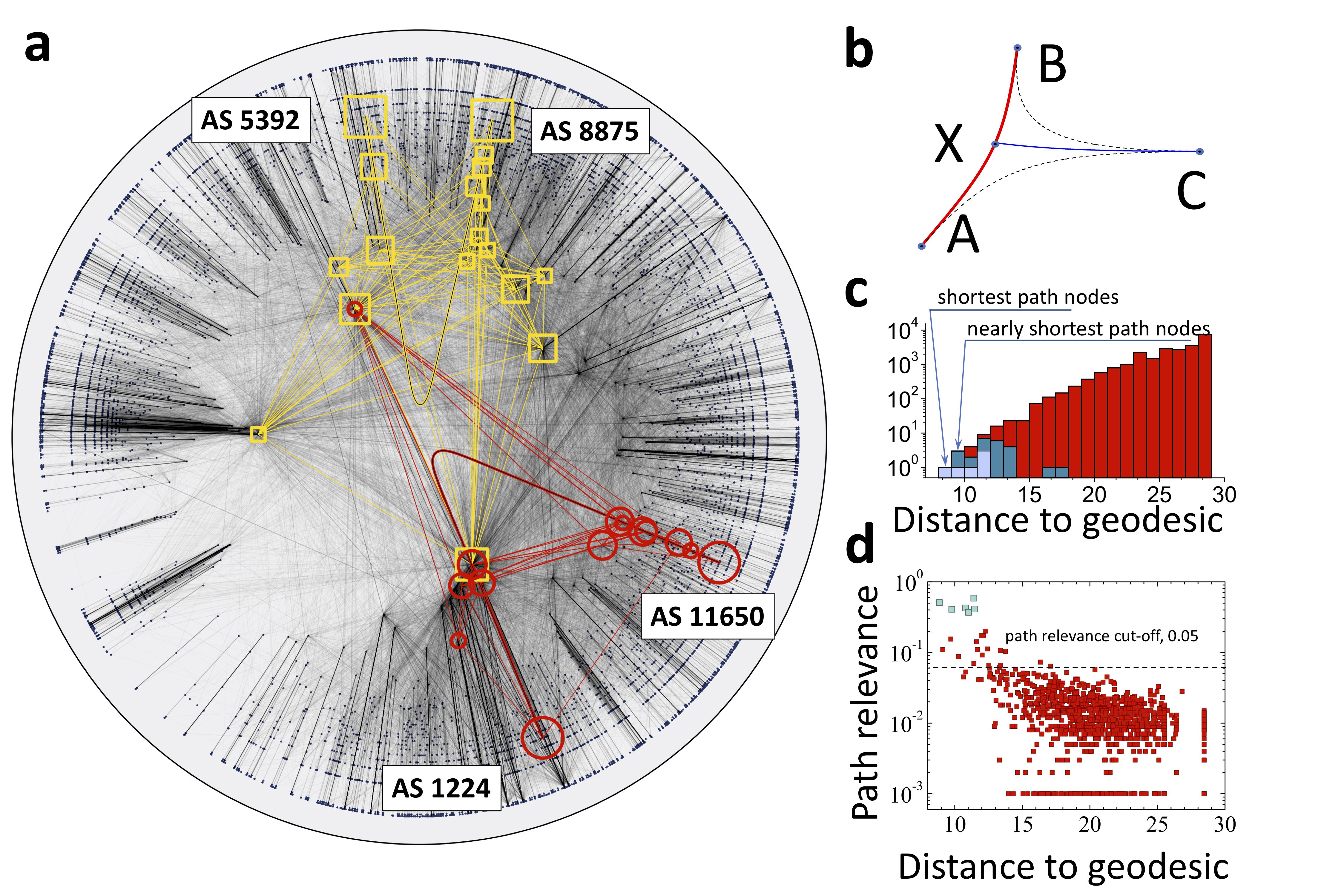}
\label{fig:1}
\end{figure}

\newpage
\clearpage
\centerline{\bf Figure 2. The accuracy of nearly shortest path finding for the AS Internet.}
\begin{figure}
\includegraphics[width=6in]{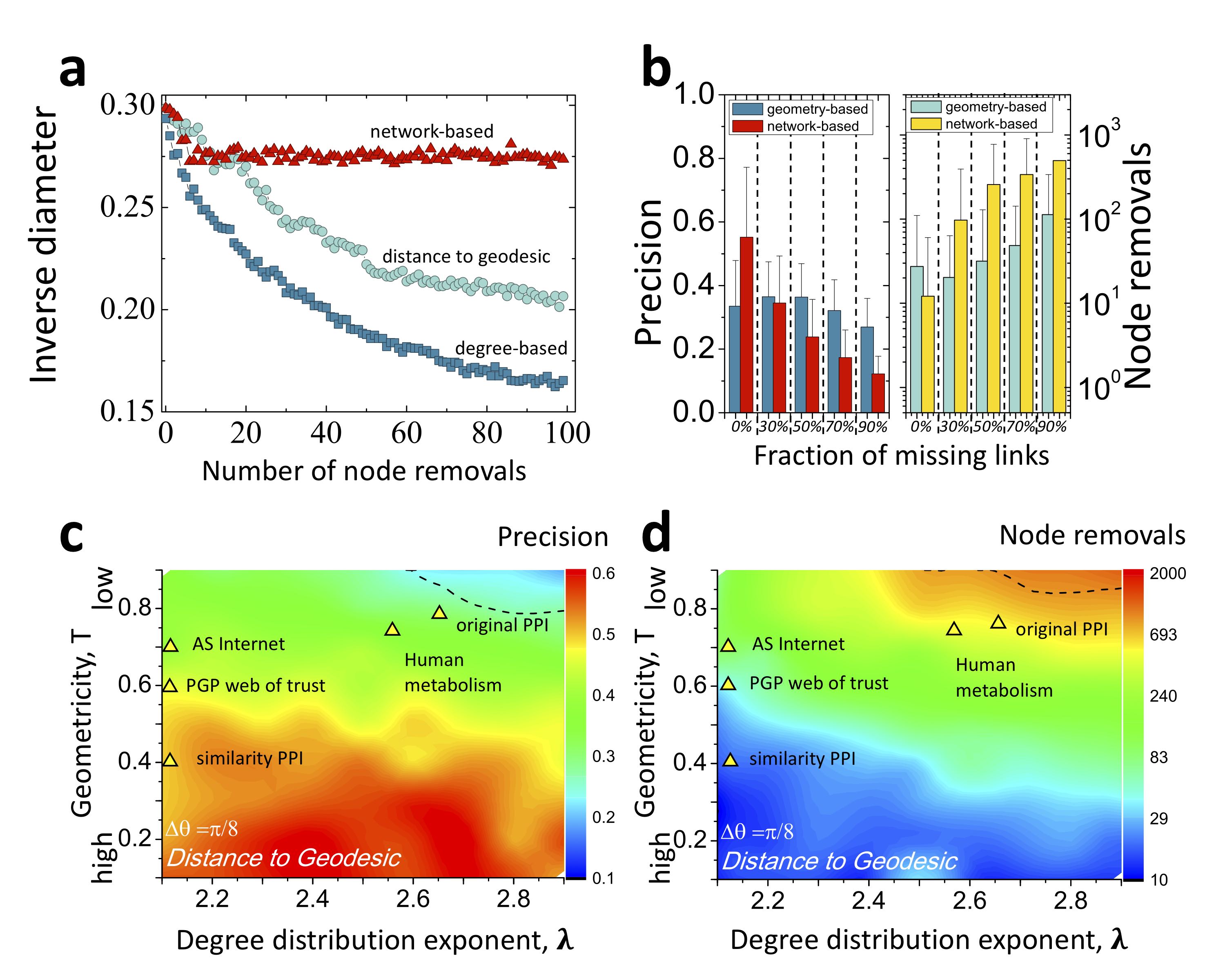}
\label{fig:2}
\end{figure}

\newpage
\clearpage
\centerline{\bf Figure 3. Anomaly detection for the interdomain Internet routing.}
\begin{figure}
\includegraphics[width=6in]{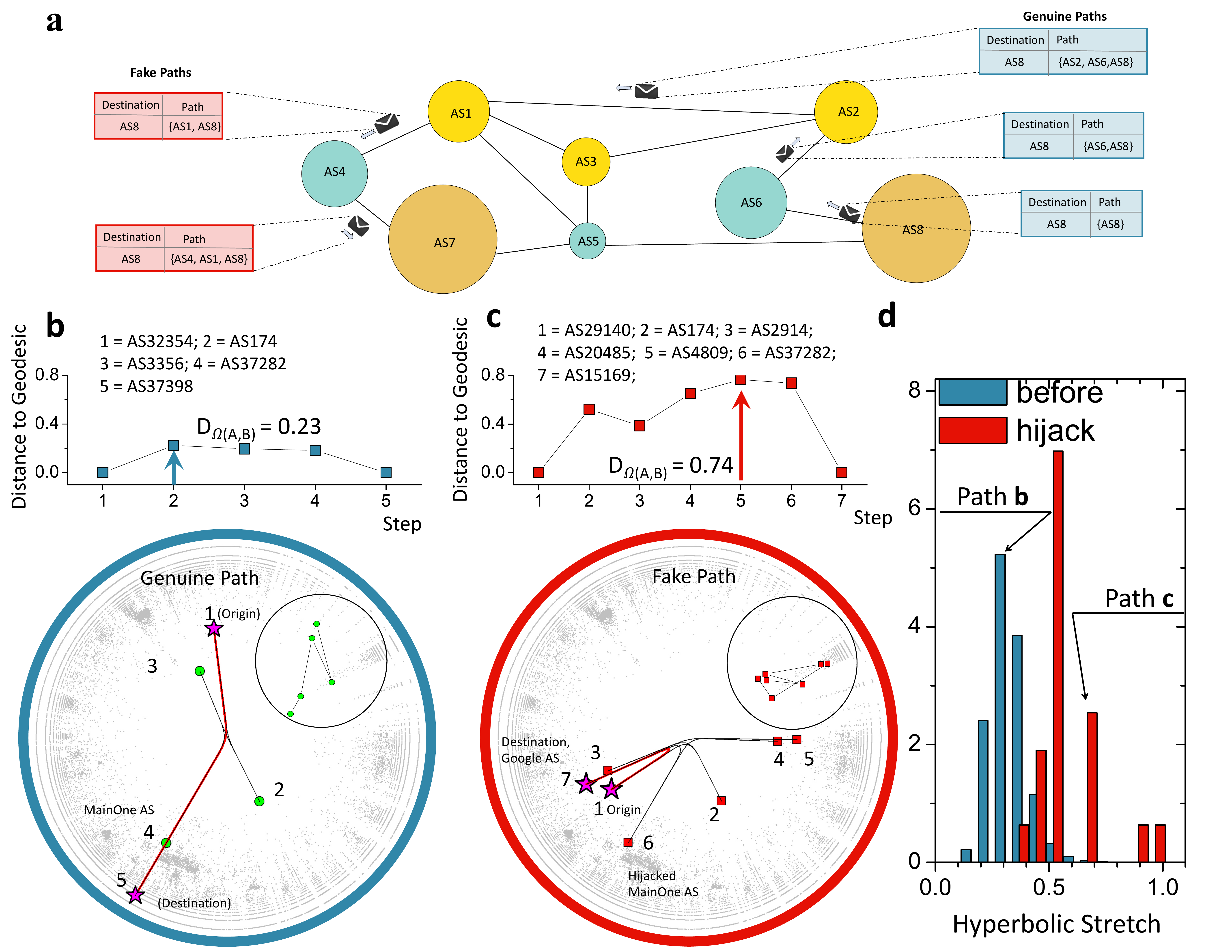}
\label{fig:3}
\end{figure}
\newpage
\clearpage
\centerline{\bf Figure 4. Latent-geometric localization of the UPP pathway.}
\begin{figure}
\includegraphics[width=6in]{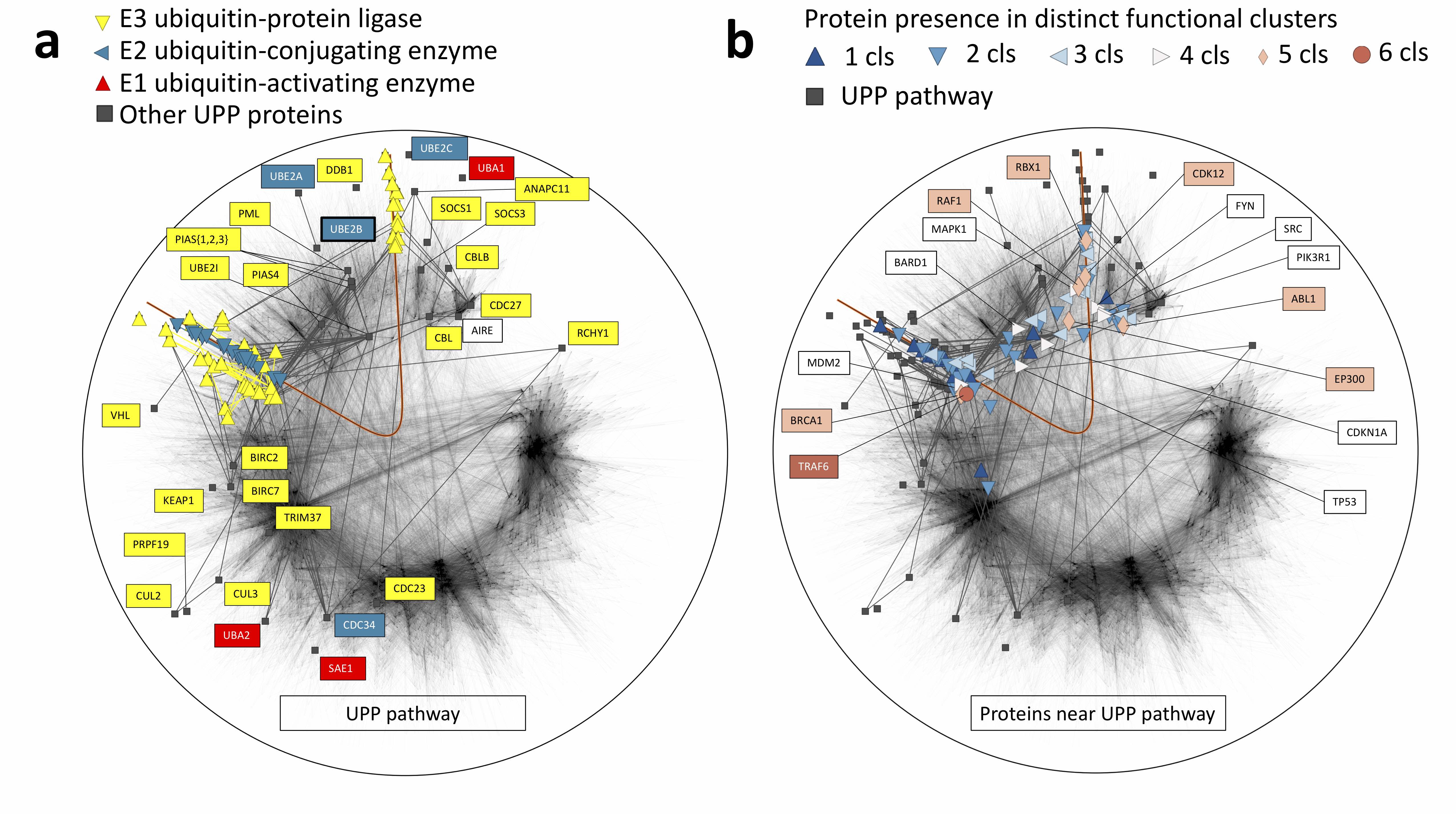}
\label{fig:4}
\end{figure}

\end{document}